# Quantifying Ultrafast Orbital Pumping and Nanometre-Scale Orbital Diffusion in Magnetic Heterostructures

Tongyang Guan[1,2,§], Jiahao Liu[1,2,§], Yuxiao Mo[1,2,§], Liangliang Zhu[3], Yizheng Wu[1,2,*], Zhensheng Tao[1,2,*]

[1] *State Key Laboratory of Surface Physics and Key Laboratory of Micro and Nano Photonic Structures (MOE), Department of Physics, Fudan University, Shanghai 200433, P R China.*

[2] *Shanghai Key Laboratory of Metasurfaces for Light Manipulation, Fudan University, Shanghai 200433, P R China.*

[3] *State Key Laboratory of Molecular Engineering of Polymers, Department of Macromolecular Science, Fudan University, Shanghai 200438, P R China.*

[§] These authors contributed equally to this work.

* Corresponding authors: wuyizheng@fudan.edu.cn; zhenshengtao@fudan.edu.cn.

## Abstract

Quantifying ultrafast orbital pumping and transport remains challenging because spin and orbital currents contribute coherently to terahertz emission. Here, we combine spintronic terahertz emission spectroscopy with wedge-sample thickness control to disentangle these contributions in Fe, Co and Ni heterostructures with Ta, W and Nb. Their opposite spin Hall and orbital Hall polarities enable quantitative determination of relative pumping strengths and propagation lengths. We find that orbital pumping increases from Fe to Ni, with inferred orbital-to-spin current ratios reaching several tens of percent in Ni, exceeding current microscopic predictions. Remarkably, Nb exhibits an orbital diffusion length of ~1.5 nm, substantially longer than in Ta and W and beyond the sub-nanometre scale predicted by recent orbital-relaxation calculations. These findings provide quantitative experimental benchmarks for ultrafast orbital pumping and transport.

## Introduction

Magnetization dynamics can inject spin angular momentum from a ferromagnet (FM) into an adjacent nonmagnetic (NM) layer through spin pumping [1–3]. Femtosecond optical excitation extends this process into the ultrafast regime, where ultrafast spin currents are launched across FM|NM interfaces, generating broadband terahertz radiation through the inverse spin Hall effect (ISHE). This enables spintronic terahertz emission spectroscopy (STES) as a contact-free probe of ultrafast angular-momentum injection and transport on picosecond timescales [4–9]. Magnetization dynamics can also generate orbital-angular-momentum (OAM) currents through orbital pumping, the orbital analogue of spin pumping [10,11]. Microscopic theory predicts that orbital-pumping efficiency relative to spin pumping increases progressively from Fe to Co to Ni [10–13], but whether this prediction applies to laser-driven ultrafast dynamics remains unclear. Meanwhile, recent orbital-relaxation calculations predict sub-nanometre orbital propagation lengths in NMs, quenched by crystal symmetry and orbital hybridization [14,15]. This conclusion is recently confirmed by experimental measurements on heavy metals [16]. Quantitative measurements of orbital-pumping efficiencies and propagation lengths in the ultrafast regime are therefore essential for establishing experimental benchmarks for the microscopic theories.

Despite recent observations of ultrafast orbital-current generation using STES [16–20], quantitative characterization remains challenging because spin and orbital currents are generated simultaneously and contribute coherently to the same terahertz emission. Previous studies have largely circumvented this difficulty by using weak spin-to-charge converters such as Cu or Ti to suppress the spin-current-induced signal and isolate the orbital channel [19,20]. Although this strategy identifies orbital contributions, it cannot quantitatively determine the spin and orbital contributions or their propagation length scales within a single experimental framework.

Here, we overcome this challenge by exploiting the distinct conversion polarities and propagation lengths of spin and orbital currents. We select Ta, W and Nb as NMs whose spin Hall and orbital Hall angles have opposite signs [21–23], producing spin- and orbital-current-induced terahertz signals with opposite polarities. Combined with high-precision wedge-sample thickness control [16], their different thickness dependences enable quantitative decomposition of the coexisting contributions in Fe, Co and Ni heterostructures.

Our results establish two quantitative benchmarks for ultrafast orbitronics. First, the orbital-pumping contribution exhibits a consistent increasing trend from Fe through Co to Ni, with the strongest response observed for Ni. Using reported NM Hall conversion factors, the inferred orbital-to-spin current-density ratios range from a few percent to ~20% for Fe and Co, whereas Ni reaches several tens of percent, exceeding current microscopic predictions. Notably, even Fe generates a non-negligible orbital contribution, revealed by near compensation between spin- and orbital-induced signals in the ultrathin-NM regime. Second, orbital diffusion is consistently shorter-ranged than spin diffusion across all NMs studied. However, Nb supports a markedly longer orbital diffusion length of ~1.5 nm, substantially exceeding those in Ta and W and the sub-nanometre scale predicted by recent orbital-relaxation calculations. These findings provide experimental benchmarks for ultrafast orbital pumping and reveal a material dependence of OAM propagation beyond current microscopic descriptions.

**Thickness-dependent terahertz emission in FM|Ta Heterostructures**

Figure 1a illustrates the general concept of our experiments. All heterostructures were grown on $Al_2O_3$(0001) substrates pre-coated with a 4-nm $SiO_2$ buffer layer. The FM layer was deposited first, followed by the NM layer, and the full heterostructure was capped with a 4-nm $SiO_2$ protective layer to prevent oxidation. The FM layer thickness was fixed at $d_{FM}$=5 nm for all Fe, Co, and Ni samples, whereas the NM layers were deposited as wedge-shaped films with thickness gradients of 0.75 - 2.5 nm/mm across a 6-mm sample length. The NM-thickness-dependent terahertz emission was measured by scanning the laser spot along the wedge-gradient direction and recording the emitted terahertz fields using free-space electro-optic sampling [24,25]. The effective thickness resolution is determined by the laser spot size (typically 100 - 200 μm in diameter), together with the local thickness gradient and scanning step size. Further details of the sample growth and experimental setup are provided in Supplementary Section S1.

Figures 1b-d compare the Ta-thickness-dependent terahertz waveforms from Fe|Ta, Ni|Ta, and Co|Ta, revealing distinct polarity evolutions. Bare 5-nm Fe, Ni, and Co films all emit terahertz signals with negative polarity with different amplitudes (see

Supplementary Section S2), providing the reference polarity for interpreting the thickness-dependent polarity evolution. These bare-FM signals can be attributed to magnetic dipole emission [26,27] and the anomalous Hall effect [28].

For Fe|Ta, Fig. 1b shows that the terahertz polarity remains negative across the measured range of Ta thickness, $d_{Ta}$=0 - 12 nm. Its polarity is opposite to that of Fe|Pt measured under the same detection geometry (see Supplementary Section S3), consistent with the opposite signs of the spin Hall angle of Ta ($\gamma_{SH}^{Ta}$) and that of Pt ($\gamma_{SH}^{Pt}$). The Fe|Ta signal is therefore dominated by ISHE-mediated spin-current conversion in Ta. Because the ISHE-induced signal and the bare-Fe contribution have the same polarity, no polarity reversal is observed in Fe|Ta as $d_{Ta}$ increases.

By contrast, Ni|Ta exhibits a polarity reversal from negative to positive at $d_{Ta}$≈0.3 nm and remains positive up to 12 nm (Fig. 1c). This positive Ta-mediated signal cannot arise from ISHE-dominated conversion. Instead, it is naturally assigned to IOHE-mediated orbital-to-charge conversion, because the Ta orbital Hall angle ($\gamma_{OH}^{Ta}$) and $\gamma_{SH}^{Ta}$ possess opposite signs [16,20].

Co|Ta exhibits a crossover behavior between Fe|Ta and Ni|Ta cases (Fig. 1d). The terahertz signal first reverses from negative to positive near $d_{Ta}$≈0.3 nm, as the orbital-current-induced emission exceeds the bare-Co emission. A second reversal occurs near $d_{Ta}$≈2.5 nm, where the negative spin-current-induced emission becomes dominant. This double polarity reversal is a particularly clear signature of competition between the orbital and spin channels in Co|Ta. The weak net emission from Co|Ta can therefore be

attributed to the near compensation between the spin and orbital components. The second reversal also indicates that the orbital contribution decays over a shorter length scale than the spin contribution, consistent with the recent work [16].

Figures 1e and f provide a direct waveform comparison that identifies the dominant contribution in Co|Ta. At large Ta thickness, Co|Ta(4) exhibits the same polarity and temporal profile as Fe|Ta(4), consistent with spin-dominated emission. At small Ta thickness, on the other hand, Co|Ta(1) closely follows Ni|Ta(2), indicating that the emission is governed by the orbital contribution. Parentheses denote the Ta thickness in nanometres.

Figure 2a presents the terahertz amplitudes extracted at $t$=0 as a function of $d_{Ta}$. Although Fe|Ta exhibits no polarity reversal, it displays a distinct threshold-like response: below $d_{Ta}$≈0.5 nm, the terahertz amplitude increases very slowly with Ta thickness, whereas such slow-growth regime is absent in Ni|Ta and Co|Ta (Fig. 2b). This behavior is unambiguously resolved using a sample containing both Ni|Ta and Fe|Ta regions that share a common Ta wedge, which thereby eliminates uncertainty in the zero-thickness position of Ta (see Supplementary Section S4).

To quantitatively disentangle the spin and orbital contributions, we adopted the multi-component terahertz-emission model (MC-TEM) following Ref. [16]. In this model, the total terahertz emission is described as the coherent sum of three contributions: the bare-FM emission, the ISHE-induced emission, and the IOHE-induced emission. Interfacial conversion processes, including the inverse spin and orbital Rashba-Edelstein effects, are excluded by the Cu- and Al-insertion control experiments (see Supplementary Section S5). The model details are provided in the Supplementary Section S6.

The MC-TEM fits reproduce the thickness-dependent amplitudes in Fig. 2a, including the absence of polarity reversal in Fe|Ta, the single reversal in Ni|Ta, and the double reversals in Co|Ta. It also captures the threshold-like behavior in Fe|Ta (Fig. 2b). Figures 2c-e further compare representative experimental and fitted waveforms for Co|Ta, confirming that the model captures not only the peak-amplitude evolution but also the temporal waveform changes. The complete set of fitting results is provided in Supplementary Section S7.

Figure 2f shows the decomposed amplitudes of spin ($A_S$), orbital ($A_L$), and bare-Fe ($A_{\mathrm{Fe}}$) contributions in Fe|Ta for $d_{\mathrm{Ta}}$<3 nm. The decomposition reveals that the threshold-like behavior originates from near cancellation between $A_S$ and $A_L$, with $A_S+A_L\approx 0$ below $d_{\mathrm{Ta}}\approx 0.5$ nm. Notably, this result shows that Fe, although generally regarded as a spin-dominated source, can induce non-negligible orbital contributions in the thin-Ta regime.

**Thickness-dependent terahertz emission FM|W Heterostructures**

We next examine the FM|W heterostructures. Figures 3a-c show the W-thickness-dependent terahertz waveforms from Fe|W, Ni|W, and Co|W, respectively. In contrast to FM|Ta, both Fe|W and Co|W maintain negative polarity over the entire W-thickness range of $d_{\mathrm{W}}$=0 - 8 nm, indicating that spin-current-induced emission dominates in these two systems. Ni|W instead exhibits two reversals (Fig. 3b): the positive orbital

component exceeds the bare-Ni signal near $d_W$≈0.2 nm, while the negative spin component regains dominance near $d_W$≈2.6 nm. The weak net emission of Ni|W relative to Fe|W and Co|W is consistent with stronger cancellation between its spin and orbital channels.

The thickness-dependent amplitudes and waveforms are well reproduced by the MC-TEM fits (Fig. 3d and Supplementary Section S8). Similar to that observed in Fe|Ta, a threshold-like behavior is present in Co|W: Co|W exhibits a nearly thickness-independent response below $d_W$≈0.6 nm, unlike Fe|W and Ni|W (Fig. 3e). The MC-TEM decomposition attributes this behavior to near cancellation between the spin and orbital contributions (Fig. 3f). This observation is corroborated by the measurements on Ni|W and Co|W fabricated on the same W wedge (see Supplementary Section S4).

**Extended orbital propagation in FM|Nb Heterostructures**

To examine whether orbital propagation remains confined to the sub-nanometre range, we turn to Nb, a lighter 4*d* transition metal with weaker SOC than the 5*d* metals Ta and W. Figures 4a-c present the terahertz waveforms as a function of Nb thickness, $d_{Nb}$, in Fe|Nb, Ni|Nb, and Co|Nb, respectively.

For Fe|Nb, the terahertz signal remains negative over the entire thickness range of $d_{Nb}$=0 - 14 nm, while Ni|Nb changes from negative to positive near $d_{Nb}$≈0.6 nm and remains orbital-dominated in the measured response at larger thicknesses (Fig. 4b). A more complex competition between spin and orbital contribution is observed in Co|Nb, where two polarity reversals appear. The first reversal occurs at $d_{Nb}$ ≈0.6 nm, where the

positive orbital signal overtakes the bare-Co emission. The second occurs at $d_{Nb}$ ≈5.4 nm, where the negative spin signal becomes dominant again at large $d_{Nb}$. Compared with Co|Ta, the second reversal is displaced to substantially larger NM thickness, immediately suggesting that the balance between the two channels extends over a longer distance in Nb.

The MC-TEM fits reproduce the Nb-thickness-dependent amplitudes and waveform evolution (Fig. 4d and Supplementary Section S9). In Fe|Nb, a threshold-like regime persists to $d_{Nb}$ ≈1.5 nm (Fig. 4e), as confirmed using Fe|Nb and Ni|Nb regions beneath the same Nb wedge (Supplementary Section S4). The decomposition again associates this behavior with near cancellation of spin and orbital contributions (Fig. 4f). Importantly, the fitted orbital diffusion length is 1.30±0.31 nm for Fe|Nb, 1.62 ±0.20 nm for Co|Nb and 1.50±0.25 nm for Ni|Nb (Supplementary Table S9). Thus, the nanometre-scale orbital propagation in Nb is reproduced across three independent FM sources.

## Discussion and Conclusion

Our measurements provide two complementary experimental benchmarks: the orbital-to-spin injection strength of the FM source and the orbital propagation length in the NM. We first consider the pumping response. Figure 5a summarizes the orbital-to-spin amplitude-coefficient ratios, $\tilde{A}_L/\tilde{A}_S$, obtained from the FM|Ta, FM|W, and FM|Nb series. These ratios characterize the relative strengths of the orbital- and spin-current-induced terahertz emission after charge conversion in the NM layer.

For a given FM layer, the extracted $\tilde{A}_L/\tilde{A}_S$ ratios are generally consistent across different NMs, indicating that they are governed by the intrinsic spin- and orbital-pumping efficiencies of the FM source layer. The sole exception is Ni|Nb. Compared to Ta and W, Nb has weaker SOC, and the spin-to-charge conversion in Ni|Nb is therefore strongly suppressed, approaching the experimental noise floor. Consequently, the fitting becomes less sensitive to the spin-induced component and tends to underestimate $\tilde{A}_S$, yielding a substantially larger apparent $\tilde{A}_L/\tilde{A}_S$ ratio with markedly increased uncertainty (Fig. 5a). Despite this caveat, $\tilde{A}_L/\tilde{A}_S$ increases systematically from Fe to Co to Ni, revealing a progressive enhancement of the orbital contribution across the 3*d* FM series, consistent with recent studies [10–12]. Notably, while Fe is commonly regarded as a dominant spin-current source, our results show that its orbital-current injection is not negligible – a finding particularly important interpreting the threshold behaviors in Fe|Ta and Fe|Nb (Figs. 2b and 4e).

For quantitative comparison with microscopic theory [10], we convert the extracted amplitude coefficients into injected current densities. According to the MC-TEM, $\tilde{A}_{L(S)} = j_{L(S)} \cdot \gamma_{\mathrm{OH(SH)}}^{\mathrm{NM}}$, where $j_{L(S)}$ is the orbital (spin) current density injected into the NM (Supplementary Section S6). The orbital-to-spin current-density ratio is therefore

$$\frac{j_L}{j_S} = \frac{\tilde{A}_L}{\tilde{A}_S} \cdot \left|\frac{\gamma_{\mathrm{SH}}^{\mathrm{NM}}}{\gamma_{\mathrm{OH}}^{\mathrm{NM}}}\right| = \frac{\tilde{A}_L}{\tilde{A}_S} \cdot \left|\frac{\sigma_{\mathrm{SH}}^{\mathrm{NM}}/\sigma_{\mathrm{C}}}{\sigma_{\mathrm{OH}}^{\mathrm{NM}}/\sigma_{\mathrm{C}}}\right|. \tag{1}$$

Here, $\sigma_{\mathrm{SH}}^{\mathrm{NM}}$ and $\sigma_{\mathrm{OH}}^{\mathrm{NM}}$ denote the NM spin Hall and orbital Hall conductivities, respectively. They are related to the corresponding Hall angles via $\gamma_{\mathrm{OH(SH)}}^{\mathrm{NM}} =$ $\sigma_{\mathrm{OH(SH)}}^{\mathrm{NM}}/\sigma_{\mathrm{C}}$, where $\sigma_{\mathrm{C}}$ is the longitudinal charge conductivity. In evaluating Eq. (1), the current-density and Hall-conversion ratios are taken as magnitudes; their signs are considered separately when assigning emission polarities.

Theoretical calculations have reported $\left|\sigma_{\mathrm{SH}}^{\mathrm{NM}}/\sigma_{\mathrm{OH}}^{\mathrm{NM}}\right|$ values that differ substantially [21–23], with values of 0.01-0.07 for Ta, 0.02-0.17 for W, and 0.01-0.04 for Nb (Supplementary Table S7). Using these reported values, we find that $j_L/j_S$ is typically in the range of a few percent to ~20% for Fe and Co, broadly consistent with microscopic theory [10]. For Ni, however, the estimated orbital-current contribution can reach several tens of percent, up to ~70% (Supplementary Table S8), indicating a substantially stronger orbital-pumping response than predicted.

Our analysis highlights a critical challenge in benchmarking orbital pumping theories: the values of $\sigma_{\mathrm{SH}}^{\mathrm{NM}}$ and $\sigma_{\mathrm{OH}}^{\mathrm{NM}}$ predicted by different ab-initio calculations differ by up to an order of magnitude for Ta, W, and Nb (Supplementary Table S7). This uncertainty propagates directly into the inferred $j_L/j_S$, whose uncertainty far exceeds the experimental error. Resolving this discrepancy requires either methodologically consistent calculations of $\sigma_{\mathrm{SH}}^{\mathrm{NM}}$ and $\sigma_{\mathrm{OH}}^{\mathrm{NM}}$ within the same theoretical framework, or direct experimental access to these quantities (e.g., via FMR spin pumping combined with orbital-current probes).

Although uncertainties in $\left|\sigma_{\mathrm{SH}}^{\mathrm{NM}}/\sigma_{\mathrm{OH}}^{\mathrm{NM}}\right|$ limit a strict quantitative comparison, the anomalously large orbital contribution in Ni remains a robust conclusion. This discrepancy may have several possible origins. First, the microscopic theory [10] considers orbital pumping driven by coherent, adiabatic magnetization dynamics, whereas STES probes angular-momentum currents generated under femtosecond-laser excitation. The ultrafast non-equilibrium dynamics may modify the relative weights of spin and orbital pumping, or activate additional orbital-current-generation channels that are not fully captured by the current theoretical model. Second, STES detects the spin and orbital angular momenta transmitted across the FM|NM interface and converted in the NM layer, rather than those generated within the FM. The interfacial transmission efficiencies of spin and orbital angular momenta may differ between the Ni|NM interfaces and their Fe|NM or Co|NM counterparts. These considerations indicate that orbital-current generation and interfacial transmission in Ni-based heterostructures require further theoretical and experimental investigations.

An additional consideration is orbital-to-spin conversion within the NM. In the presence of SOC, orbital currents can generate secondary spin currents that may also produce terahertz emission through ISHE [29–32]. This process is not explicitly included in the present MC-TEM, where the positive-polarity charge current is solely attributed to direct IOHE-mediated orbital-to-charge conversion. Because the orbital-to-spin conversion coefficient is negative for Ta, W, and Nb [17,23], the secondary spin currents would produce a terahertz signal with the same positive polarity as the IOHE contribution. If this secondary pathway were significant, the apparent $\tilde{A}_L/\tilde{A}_S$ ratio should show a clear systematic dependence on the NM layer. In particular, the weaker SOC of Nb should suppress orbital-to-spin conversion and make the Nb-derived ratio markedly different from those obtained with Ta and W for the same FM source layer. However, such a systematic deviation is not observed in Fig. 5a. Excluding Ni|Nb, where the strongly suppressed spin-to-charge conversion leads to an overestimated apparent $\tilde{A}_L/\tilde{A}_S$, the Nb-derived ratios remain comparable to those obtained from Ta and W in Fe- and Co-based heterostructures. We therefore conclude that the secondary orbital-to-spin conversion is unlikely to play a significant role in the present analysis.

Our second benchmark concerns the propagation of OAM in NMs. Across all heterostructures, the extracted orbital diffusion length, $\lambda_L$, is consistently shorter than the corresponding spin diffusion length, $\lambda_S$ (Fig. 5b and Supplementary Table S9). Strikingly, $\lambda_L$ is markedly longer in Nb (~1.5 nm) than those in the heavier 5*d* metals Ta and W (0.3 - 0.8 nm). This extended $\lambda_L$ in Nb raises two key implications. First, it indicates that OAM can indeed transport deep into Nb, giving rise to an orbital current, whereas in Ta and W it remains localized near the interface as orbital polarization [16]. Second, the pronounced material dependence of $\lambda_L$ suggests that orbital propagation may be influenced by the SOC strength and crystal-field environment of the NM. Notably, the nanometre-scale orbital diffusion length in Nb exceeds the sub-nanometre scale predicted by recent microscopic orbital-relaxation calculations for several other transition metals [14]. Understanding this discrepancy will require microscopic descriptions that account for orbital relaxation, crystal-field effects and interfacial transmission.

In conclusion, by resolving the overlapping spin- and orbital-current signals in time-domain terahertz emission, we have obtained quantitative benchmarks for both ultrafast orbital injection and its propagation. The enhanced Ni response and nanometre-scale orbital transport in Nb expose limits of existing near-equilibrium pumping and orbital-relaxation descriptions, motivating microscopic treatments of nonequilibrium dynamics, conversion and interfaces. More broadly, the approach provides a route towards materials-specific control of spin and orbital angular-momentum flow.

**Acknowledgments:** This work was accomplished in Fudan University. We acknowledge the support from the National Key Research and Development Program of China (Grant Nos. 2021YFA1400200, 2024FYA1408500, 2022YFA1403300 and 2022YFA1404700). Z.T. and Y. W. acknowledge the support from the National Natural Science Foundation of China (Grant No. 12221004). Z. T. acknowledges the support from the National Natural Science Foundation of China (Grant Nos. 12450407 and 12274091). Y. W. acknowledges the support from the National Natural Science Foundation of China (Grant Nos. 12274083, 12434003,12221004).

**Data availability:** The minimum dataset is available in the main text, the supplementary information and the supplementary dataset. Additional data are available upon request to the corresponding author, Y. W. and Z. T.

**Author Contributions Statement:**

Conceptualization: Y. W., Z. T.

STES Experiment: T. G., Y. M.

Sample preparation: J. L.

Supervision: Y. W., Z. T., L. Z.

Writing – original draft: T. G., Z. T., Y. W.

Writing – review & editing: J. L., Y. M., L. Z.

**Competing Interests Statement:** Authors declare that they have no competing interests.

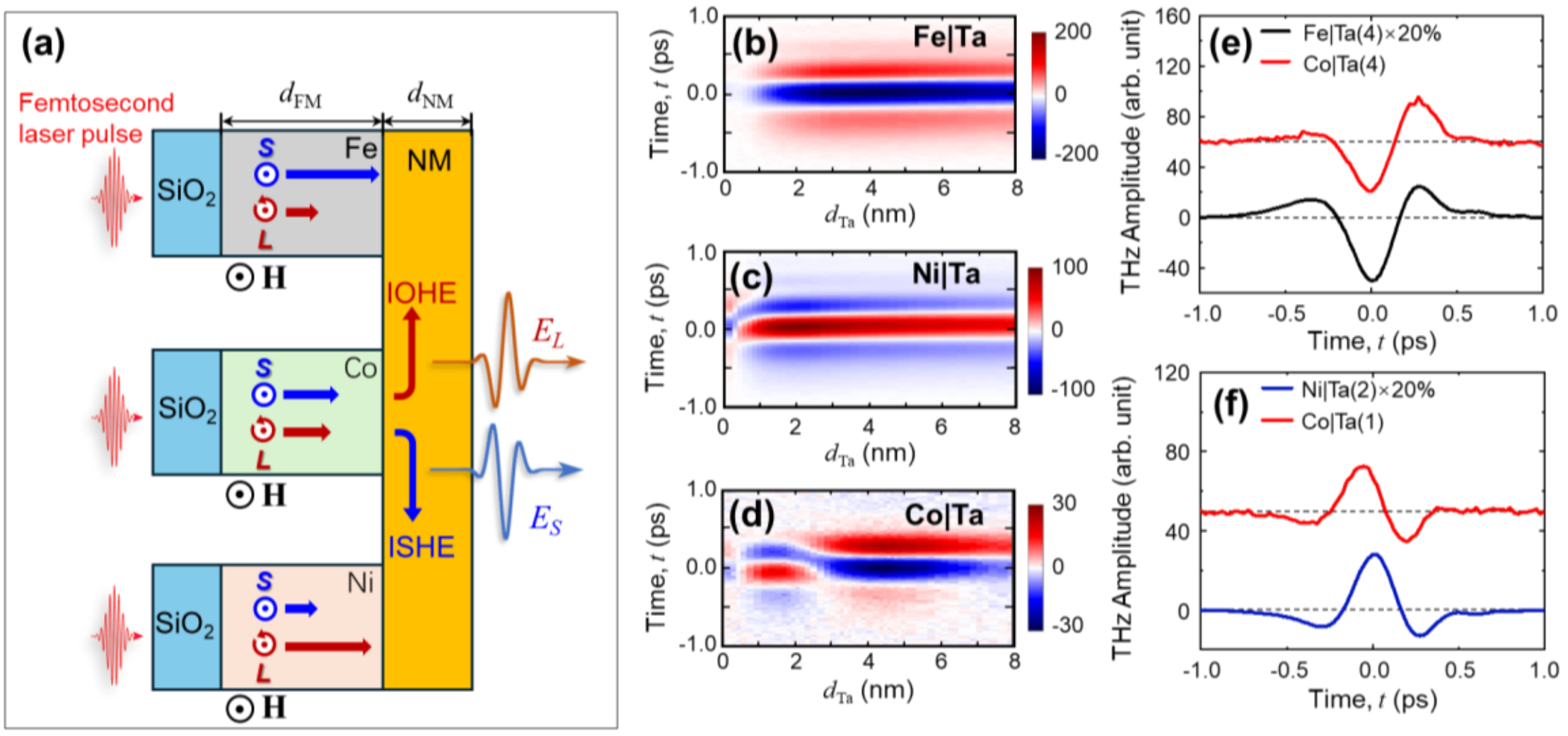

**Figure1. Experimental scheme and terahertz emission from the FM|Ta heterostructures. a.** Schematics of the STES experiments on FM|NM heterostructures. **b.- d.** Two-dimensional plots of terahertz waves at different Ta thickness, $d_{Ta}$, for Fe|Ta, Ni|Ta, and Co|Ta, respectively. **e.** Representative terahertz waveforms from Fe|Ta(4) and Co|Ta(4). For clarity, the Co|Ta(4) waveform is vertically offset by +50, and the amplitude of Fe|Ta(4) is scaled to 20% of its original value. **f.** Representative terahertz waveforms from Ni|Ta(2) and Co|Ta(1). For clarity, the Co|Ta(1) waveform is vertically offset by +50, and the amplitude of Ni|Ta(2) is scaled to 20% of its original value. Numbers in parentheses denote the Ta thickness in nanometres.

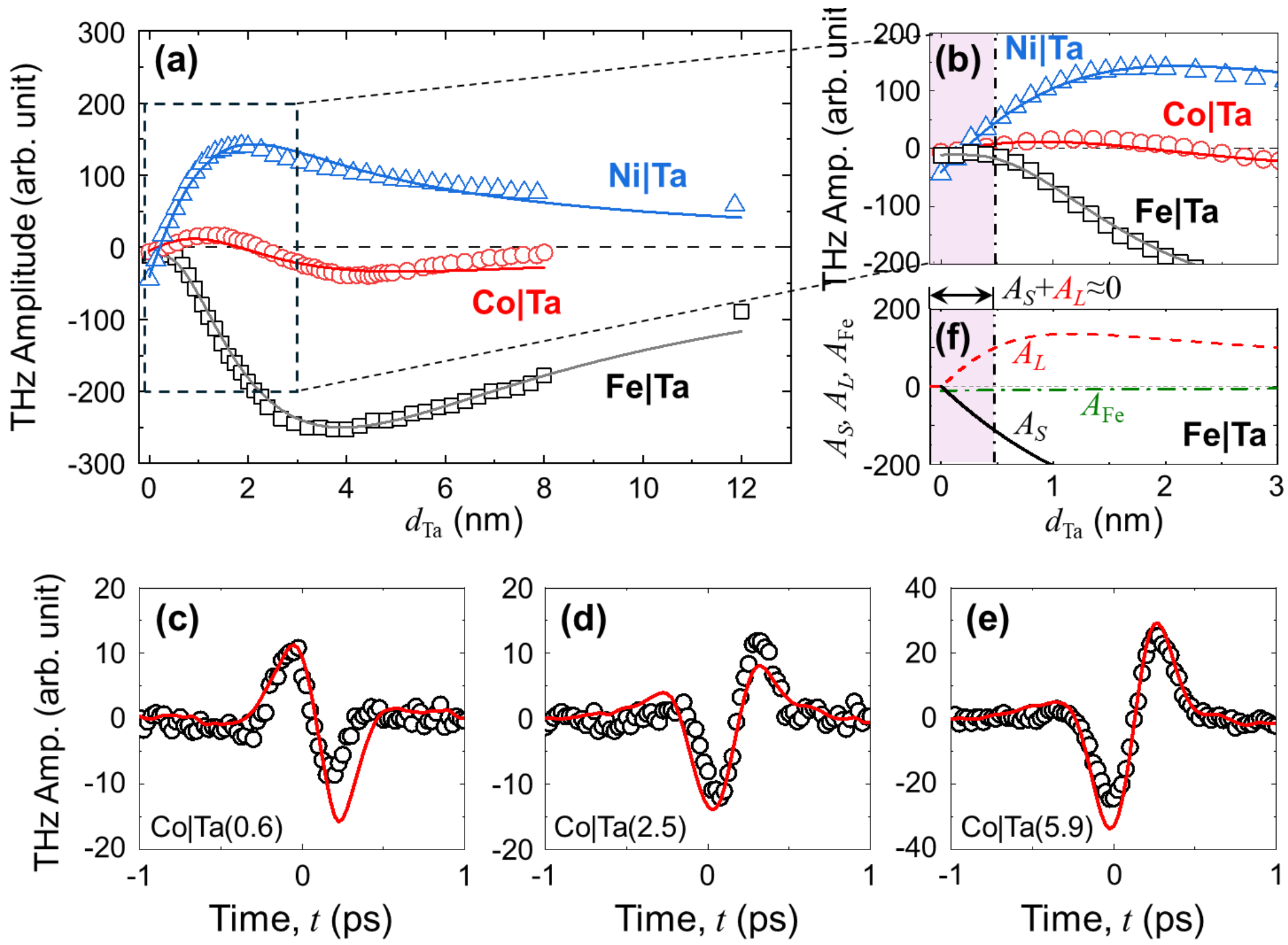

**Figure2. MC-TEM decomposition of terahertz signals from the FM|Ta heterostructures. a.** Terahertz amplitudes extracted at $t$=0 as a function of Ta thickness, $d_{\mathrm{Ta}}$, for the Fe|Ta, Co|Ta, and Ni|Ta heterostructures. Symbols represent experimental data; solid lines indicate fits based on the MC-TEM. **b.** Zoom-in view of **a.** for $d_{\mathrm{Ta}}$ <3 nm. The threshold-like regime in Fe|Ta is highlighted by the grey shaded region and vertical dash-dotted line. **c.- e.** Representative MC-TEM fits to time-domain terahertz waveforms from Co|Ta samples with different $d_{\mathrm{Ta}}$. Symbols represent experimental data; solid lines show model fits. **f.** Extracted spin ($A_S$), orbital ($A_L$), and bare-Fe ($A_{\mathrm{Fe}}$) contributions in Fe|Ta as a function of $d_{\mathrm{Ta}}$. The grey shaded region corresponds to the threshold-like regime shown in **b.**, where the spin and orbital contributions nearly compensate with each other, i.e. $A_S + A_L \approx 0$.

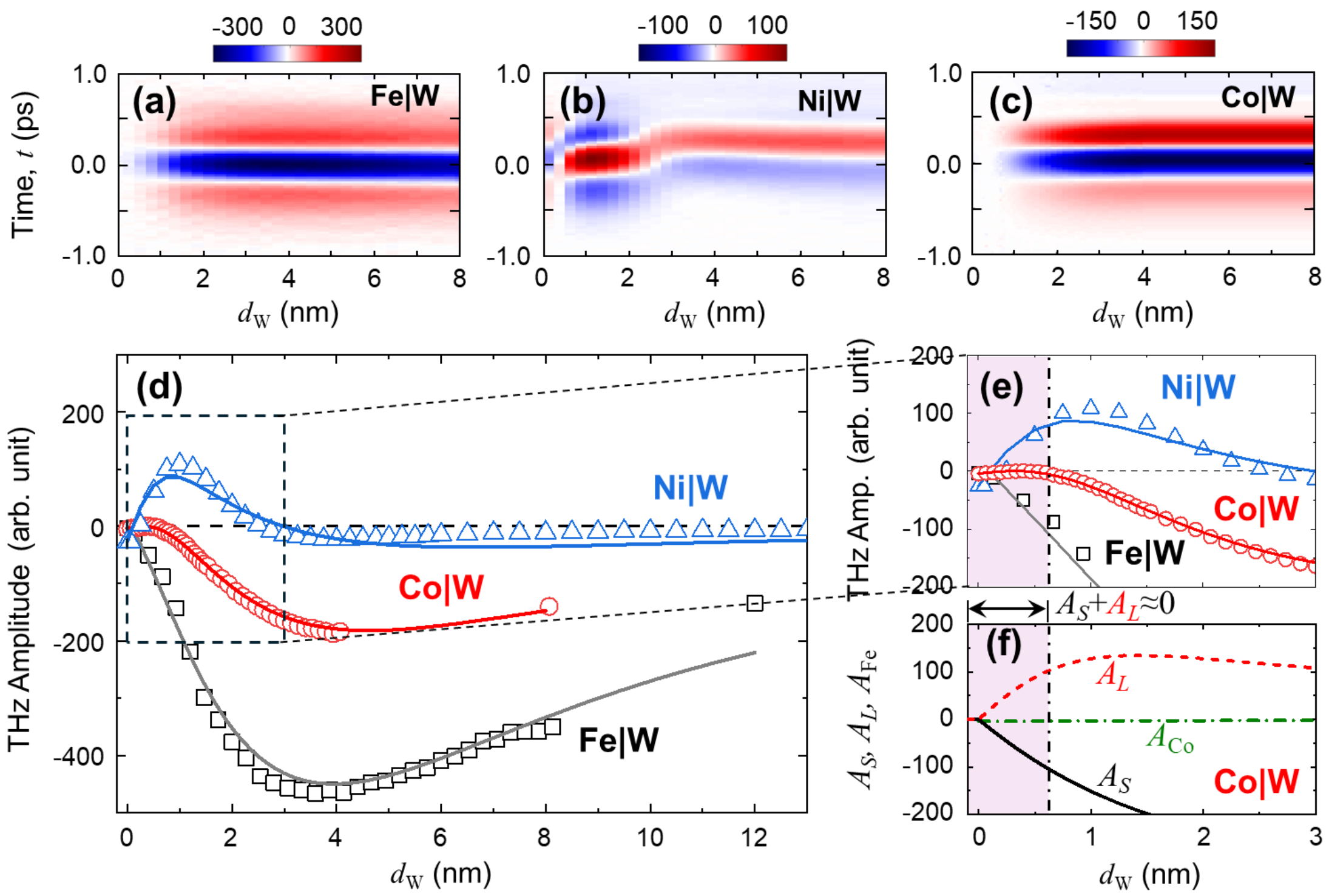

**Figure3. Terahertz emission from the FM|W heterostructures. a.-c.** Two-dimensional plots of terahertz waves at different W thickness, $d_W$, for Fe|W, Ni|W, and Co|W, respectively. **d.** Terahertz amplitudes extracted at $t$=0 as a function of $d_W$, for the FM|W heterostructures. Symbols represent experimental data; solid lines indicate fits based on the MC-TEM. **e.** Zoom-in view of **d.** for $d_W$ <3 nm. The threshold-like regime in Co|W is highlighted by the grey shaded region and vertical dash-dotted line. **f.** Extracted $A_S$, $A_L$, and $A_{Co}$ contributions in Co|W as a function of $d_W$. The grey shaded region corresponds to the threshold-like regime shown in **e.**, where the spin and orbital contributions nearly compensate with each other.

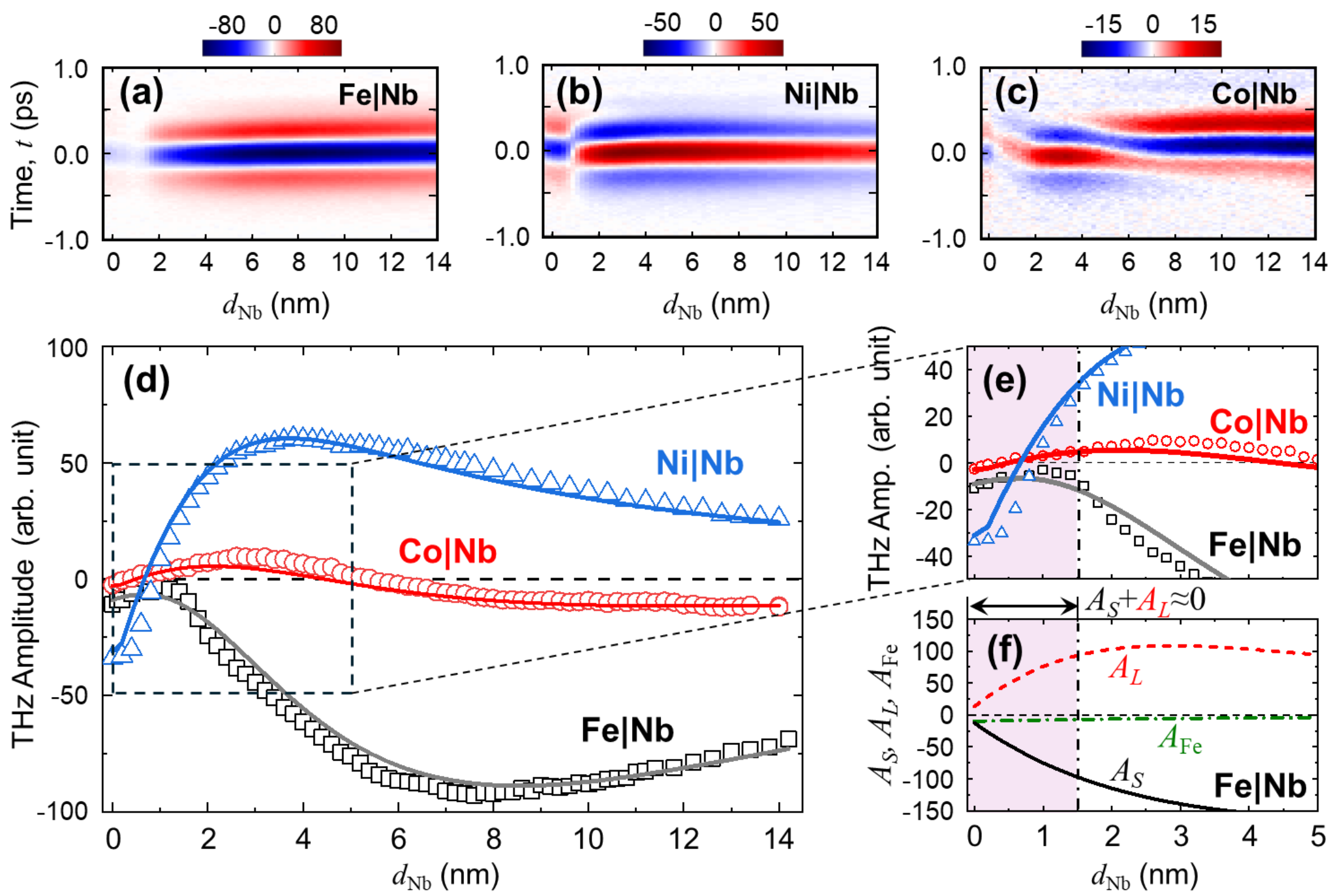

**Figure4. Terahertz emission from FM|Nb heterostructures. a.-c.** Two-dimensional plots of terahertz waves at different Nb thickness, $d_{Nb}$, for Fe|Nb, Ni|Nb, and Co|Nb, respectively. **d.** Terahertz amplitudes extracted at $t$=0 as a function of $d_{Nb}$, for FM|Nb heterostructures. Symbols represent experimental data; solid lines indicate fits based on the MC-TEM. **e.** Zoom-in view of **d.** for $d_{Nb}$ <5 nm. The threshold-like regime in Fe|Nb is highlighted by the grey shaded region and vertical dash-dotted line. **f.** Extracted $A_S$, $A_L$, and $A_{Fe}$ contributions in Fe|Nb as a function of $d_{Nb}$. The grey shaded region corresponds to the threshold-like regime shown in **e.**, where the spin and orbital contributions nearly compensate with each other.

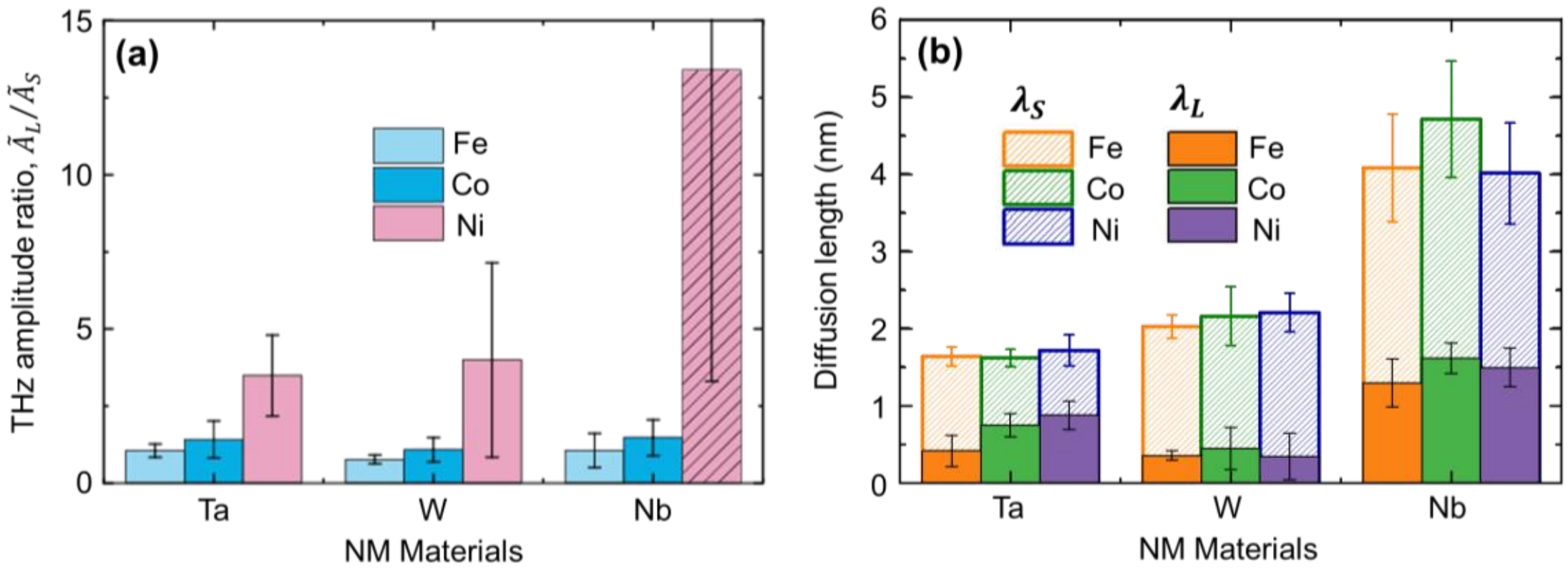

**Figure 5. Quantitative characterization of ultrafast spin and orbital pumping and transport. (a)** Orbital-to-spin terahertz amplitude-coefficient ratios, $\tilde{A}_L/\tilde{A}_S$ , for Fe-, Co-, and Ni-based heterostructures with Ta, W, and Nb. The Ni|Nb bar is hatched to indicate its large uncertainty arising from the weakly constrained spin contribution. **(b)** Spin and orbital diffusion lengths ($\lambda_S$ and $\lambda_L$) in Ta, W, and Nb with different FM source layers. All values are extracted from fits to the experimental data using the multi-component terahertz-emission model (MC-TEM). Error bars indicate fitting uncertainties.